\documentclass[prb,amsmath,twocolumn,letterpaper]{revtex4-2}
\usepackage{graphicx}
\usepackage{amssymb}
\begin{document}

\title{Doping asymmetry in the three-band Hamiltonian for cuprate ladders:
failure of the standard model of superconductivity in cuprates}
\author{Jeong-Pil Song}
\affiliation{Department of Physics, The University of Arizona Tucson, AZ 85721}
\author{Sumit Mazumdar}
\affiliation{Department of Physics, The University of Arizona Tucson, AZ 85721}
\author{R. Torsten Clay}
\affiliation{Department of Physics \& Astronomy, and HPC$^2$ Center for Computational Sciences, Mississippi State University, Mississippi State, MS 39762}
\date{today}
\begin{abstract}
The relevance of the single-band two-dimensional Hubbard model to
superconductivity in the doped cuprates has recently been questioned,
based on Density matrix Renormalization Group (DMRG) computations that
found superconductivity over unrealistically broad doping region upon
electron-doping, yet complete absence of superconductivity for
hole-doping. We report very similar results from DMRG calculations on
Cu$_2$O$_3$ two-leg ladder within the parent three-band
correlated-electron Hamiltonian. The strong asymmetry found in our
calculations are in contradiction to the deep and profound symmetry in
the experimental phase diagrams of electron- and hole-doped cuprate
superconductors, as seen from the occurrence of quantum critical
points within the superconducting domes in both cases that are
characterized by Fermi surface reconstruction, large jumps in carrier
density and strange metal behavior.
\end{abstract}
\maketitle

The mechanism of unconventional superconductivity (SC) found in the
high-T$_c$ cuprates and other strongly-correlated materials remains an
outstanding problem in condensed matter physics, more than three
decades after its discovery. At the heart of the problem is the choice
of the minimal model for the CuO$_2$ planes that can account for SC.
Since the work of Zhang and Rice, who showed that under certain limits
the three-band model of the CuO$_2$ planes could be reduced to a
simpler one-band Hubbard model \cite{Zhang88a}, the majority of
theoretical work has focused on the single-band Hubbard model, as well
as even simpler approximations such as the $t$-$J$ model. While
cluster variants of dynamical mean-field theory find SC in the doped
single-band model on a square lattice
\cite{Lichtenstein00a,Maier00a,Senechal05a,Capone06a,Kancharla08a,Gull13a,Foley19a,Kitatani22a},
density matrix renormalization group (DMRG) and quantum Monte Carlo
(QMC) calculations have detected absence of long-range superconducting
order \cite{Zhang97b,Qin20a,Vaezi21a}.

Accurate description of the band structure of the cuprates within a
one-band correlated-electron Hamiltonian requires inclusion of second
neighbor hopping $t^\prime$
\cite{Hybertsen90a,Muller98a,Hirayama18a,Hirayama19a}.  DMRG
calculations have therefore been performed on quasi-one-dimensional
cylinders for the $t$-$t^\prime$-$J$ model, where $t^\prime/t$
negative (positive) corresponds to hole (electron)-doped regimes.  No
signature of pairing is found in the negative $t^\prime/t$ region
\cite{Jiang21a}. Surprisingly, strong signature of dominant
superconducting pair-pair correlations is found in the positive
$t^\prime/t$ region, over a very broad range of electron-doping
\cite{Jiang21a}. Enhanced pairing correlations in the electron-doped
region have been confirmed from DMRG calculations on related extended
$t$-$J$ models on one-band 6-leg cylinders
\cite{Gong21a,Jiang21b}. These results are exactly opposite to
experimental observation in real cuprates, where significantly higher
T$_c$ over a much broader doping region is found with hole doping.
The authors of reference \onlinecite{Jiang21a} have subsequently
extended their calculations to the parameter region with nonzero third
neighbor hopping $t^{\prime\prime}$ \cite{Jiang21b}. Absence of
pairing in the hole-doped region, and strong pairing tendency over
very broad region of electron-doping persist within the
$t$-$t^\prime$-$t^{\prime\prime}$-$J$ model \cite{Jiang22a}.  
Quantum Monte Carlo calculations have claimed long-range
superconducting correlations for both electron and hole doping at
finite $U$ in the $U$-$t$-$t^\prime$ Hamiltonian, with {\it stronger}
pairing on the hole doped side \cite{Xu23a}.  DMRG calculations
for the same model contradict these results, however, and only find
pairing on the electron-doped side \cite{Jiang23a}.  The origin of the
differences in these numerical results and the more serious
discrepancy from experimental observations remain not understood.

The single-band model calculations suggest that there are potential
problems with reducing the electronic structure of the CuO$_2$ planes
to Cu-site based effective models.  Clearly a comparison of hole-
versus electron-doped pairing tendencies within the full three-band
correlated-electron Hamiltonian for the cuprates will be more useful
in this context.  We report here the results of high precision DMRG
computations on the three-band two-leg cuprate ladder, over a wide
range of hole- and electron-doping.  The corresponding single-band
Hubbard ladder has been widely investigated in the past
\cite{Noack94a,Noack97a,Balents96a,Dolfi15a,Gannot20a}.  The undoped
(half-filled) single-band two-leg Hubbard ladder has spin-gapped
ground state, with spins on the ladder rungs paired into singlets
\cite{Dagotto96a}. Doped holes or electrons (equivalent since the
single-band Hubbard ladder has particle-hole symmetry) occupy ladder
rungs in pairs, which is favored over unpaired charge carriers that
would destroy two singlets instead of one.  The ground state of the
single-band ladder for weak to moderate doping consequently has a spin
gap and exhibits singlet superconducting correlations with
quasi-long-range order
\cite{Noack94a,Noack97a,Balents96a,Hur09a,Dolfi15a,Gannot20a}.  The
above result breaks down for the hole-doped three-band ladder, where a
recent DMRG study has shown that even though a spin gap persists in
the undoped state, superconducting correlations in the hole-doped
decay {\it faster} than $1/r$, indicating dominance of charge over
pairing correlations at long distances \cite{Song21a}. The decay of
pair correlations in this case is caused by pair-breaking hole hopping
between the O ions, and is strongest when both Coulomb interactions
between holes on the same O and O-O hopping are included
\cite{Song21a}. The doped holes in hole-doped cuprates primarily
reside on oxygen sites; the results for the hole-doped ladder indicate
a breakdown of the Zhang-Rice theory \cite{Zhang88a}. In what follows
we compare hole- versus electron-doped three-band two-leg ladder
within high precision calculations.

We consider the Cu$_2$O$_3$ two-leg ladder Hamiltonian,
\begin{align}
&H = \Delta_{\rm dp}\sum_{i\sigma} p^\dagger_{i,\sigma}p_{i,\sigma}
 +\sum_{\langle ij \rangle, \lambda, \sigma}t_{\rm dp}^{\perp}
     (d^\dagger_{i,\lambda,\sigma}p_{j,\sigma}+H.c.)\nonumber \\
&+\sum_{\langle ij \rangle,\lambda, \sigma}t_{\rm dp}(d^{\dagger,i,\sigma}_{i,\lambda,\sigma}p_{j,\sigma}+H.c.)
     +\sum_{\langle ij \rangle, \sigma}t_{\rm pp} 
     (p^\dagger_{i,\sigma}p_{j,\sigma}+H.c.)\nonumber \\
 &+U_{\rm d}\sum_{i,\lambda} d^\dagger_{i,\lambda,\uparrow}d_{i,\lambda,\uparrow}d^\dagger_{i,\lambda,\downarrow}d_{i,\lambda,\downarrow}
     +U_{\rm p}\sum_j p^\dagger_{j,\uparrow}p_{j,\uparrow}p^\dagger_{j,\downarrow}p_{j,\downarrow}
     \label{hamiltonian}
\end{align}

In Eq.~\ref{hamiltonian} $d^\dagger_{i,\lambda,\sigma}$ creates a hole
with spin $\sigma$ on the $i$th Cu-site on the $\lambda$-th leg
($\lambda$=1,2) of the ladder, $p^\dagger_{j,\sigma}$ creates a hole
of spin $\sigma$ on the $j$-th O $p$ orbital. The O-ion can be located
on a rung or either leg of the ladder.  Parameters $t^{\perp}_{\rm
  dp}$ and $t_{\rm dp}$ are the nearest neighbor (n.n.)  Cu-O rung and
leg hopping integrals, respectively, while $t_{\rm pp}$ is the
n.n. O-O hopping integral. The phase relations between the orbitals
(see Fig.~S1 in Supplemental Material) determine the sign
convention for the hopping integrals. We have taken all
$t^{\perp}_{\rm dp}$ as negative, while $t_{\rm dp}$ and $t_{\rm pp}$
alternate signs along the length of the ladder.  $U_{\rm d}$ ($U_{\rm
  p}$) is the Hubbard repulsion between hole pairs on Cu-$d$ (O-$p$)
orbitals, and $\Delta_{\rm dp}=\epsilon_{\rm p}-\epsilon_{\rm d}$ is
the site-energy difference between Cu-$d$ and O-$p$ orbitals.  We
consider ladders with $L$ rungs and open boundary condition, with
rungs at both terminal ends. Calculations are for ladders up to $L=96$
(192 Cu and 286 O sites) and $N$ holes, with the undoped state
corresponding to one hole per Cu site ($N=2L$).  For hole (electron)
doping we add (remove) particles and define the hole (electron) doping
fraction as $\delta_{\rm h}$ ($\delta_{\rm e}$) = $N/(2L)-1$
($1-N/(2L)$).  In the following we make comparisons of three-band
results with those obtained from single-band Hubbard ladders.  The
single-band Hubbard repulsion, and the rung and leg hopping parameters
are written as $U$, $t$ and $t^{\perp}$, respectively. The single-band
doping fraction is written as $\delta$.

\begin{figure}[tb]
  \centerline{\resizebox{3.5in}{!}{\includegraphics{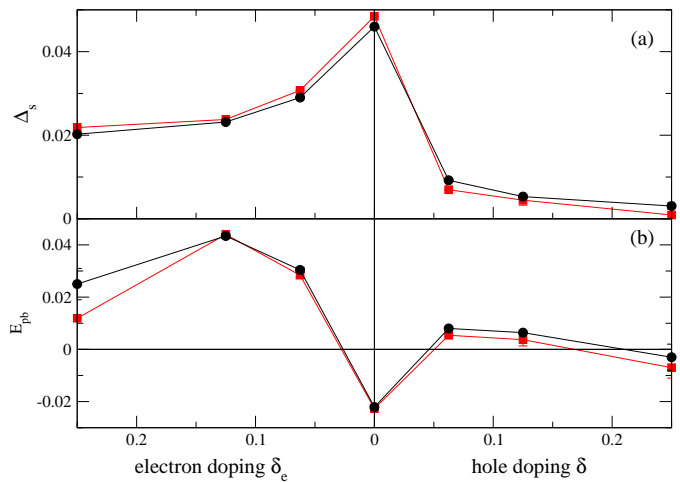}}}
  \caption{(Color online) (a) The doping dependence of the
    extrapolated spin gaps $\Delta_s$ in the infinite-length limit
    ($L\rightarrow\infty$). (b) Pair-binding energy E$_{\rm pb}$ as a
    function of doping (see text).  Circles and squares are for
    $(U_{\rm p},t_{\rm pp})=(3,0.5)$ and $(4,0.6)$, respectively. A
    transition to a band state with near-equal populations of
    charge-carriers on Cu- and O-sites occurs at $\delta_e$ larger
    than that shown here.  Lines are guides to the eye.}
  \label{fig1}
\end{figure}

We set $|t_{\rm dp}|$=1 ($t^{\perp}_{\rm dp}=-1$) and take other
Hamiltonian parameters from recent first-principles calculations,
$\Delta_{\rm dp}=3$, $U_d=8$, $U_p=\{3,4\}$, and $t_{\rm
  pp}=\{0.5,0.6\}$ \cite{Hirayama18a,Hirayama19a}.  These parameters
are similar to commonly accepted values
\cite{Jeckelmann98a,Nishimoto02a,White15a}.  We employed an
$S_z$-conserving DMRG algorithm using the ITensor library
\cite{itensor} with real-space parallelization \cite{Stoudenmire13a}.
We used a maximal bond dimension of up to 19000, giving a truncation
error of less than 1$\times$10$^{-7}$.  All results were extrapolated
to the limit of zero truncation error (see \cite{Song21a} for examples
of extrapolation).

The characteristic behavior of the two-leg ladder is determined by its
spin gap $\Delta_s$.  SC can occur only if the spin gap found in the
undoped ladder persists under doping \cite{Hur09a,Dagotto92a,Dolfi15a}. We calculated $\Delta_s$ using
finite-size extrapolation from ladders of lengths up to $L$=64.
Fig.~\ref{fig1}(a) shows the doping dependence of the
$L\rightarrow\infty$ extrapolated $\Delta_s$.  For the undoped ladder,
the behavior of $\Delta_s$ against $U_d/|t_{\rm pd}|$ is very similar
to that of the spin gap versus $U/t$ in the single-band Hubbard ladder
\cite{Noack94a}, with a maximum in $\Delta_s$ for $U_d/|t_{\rm
  pd}|\approx$ 8 \cite{Song21a}.  However, $\Delta_s$ behaves {\it
  qualitatively differently} for the electron versus hole doped
ladders within Eq.~\ref{hamiltonian}.  For electron doped ladders
$\Delta_s$ remains large over a wide doping range, while for hole
doping $\Delta_s$ decreases rapidly with doping.  The normalized spin
gap $\tilde{\Delta}_s\equiv\Delta_s(\delta_e)/\Delta_s(\delta_e=0)$
for the electron doped ladder is comparable to $\tilde{\Delta}_s$ for
the single-band Hubbard ladder with $U$=8 and $t^\perp=t$
\cite{Noack96a}.  For the single-band ladder,
$\tilde{\Delta}_s(\delta=0.125)\approx$ 0.42, and is only slightly
smaller at $\delta$ = 0.25 \cite{Noack96a}; in comparison, for the
electron doped cuprate ladder with $U_{\rm d}=8$, $U_{\rm p}=3$, and
$t_{\rm pp}=0.5$, $\tilde{\Delta}_s(\delta_e=0.125)$ = 0.49 and
$\tilde{\Delta}_s(\delta_e=0.25)$ = 0.45.  However, for hole doping,
$\tilde{\Delta}_s(\delta_h=0.125)$ = 0.14 and
$\tilde{\Delta}_s(\delta_h=0.25)$ = 0.02.  $\Delta_s$ increases with
increasing $t_{\rm pp}$ in the undoped three-band model
\cite{Nishimoto02a}. This effect can be explained in the undoped case
from perturbative calculations of the effective exchange $J$ between
n.n. Cu spins.  About 2/3 of the contribution to $J$ involves $t_{\rm
  pp}$, demonstrating the critical role that the oxygen sublattice
plays even in undoped cuprates \cite{Eskes93a}.  Our DMRG results show
that while $\Delta_s$ increases with $t_{\rm pp}$ with electron
doping, $\Delta_s$ {\it decreases} with $t_{\rm pp}$ for hole doping.
We also calculated the finite-size scaled
pair-binding energy E$_{\rm pb}$ for both hole- and electron-doping, defined as in \cite{Abdelwahab23a},
\begin{equation}
  E_{\rm pb}=2E(N_\uparrow-1,N_\downarrow)-E(N_\uparrow-1,N_\downarrow-1)-E(N_\uparrow,
  N_\downarrow).
  \label{PB}
\end{equation}
The calculated pair-binding energies, shown in Fig.~1(b), are consistent with the calculated $\Delta_s$.

The doped single band two leg ladder belongs to the Luther-Emery
universality class, with gapped spin degrees of freedom and a single
gapless charge mode \cite{Luther74a,Balents96a,Dolfi15a,Hur09a}.  For
the three-band cuprate ladder we define the local charge density
operator $n_j$ for the $j$th unit cell as the sum of the charge
density operators for the two Cu sites on a rung, the rung O, and two
leg O sites. The charge correlation function is defined as
$C(r)=\langle n_i n_j -\langle n_i \rangle \langle n_j
\rangle\rangle$, where $r\equiv|i-j|$ is the rung-rung distance.  We
define the superconducting pair-pair correlation function
$P(r)=\frac{1}{2}(\langle \Delta_i^\dagger \Delta_j\rangle + \langle
\Delta_i \Delta_j^\dagger \rangle)$, where
$\Delta_i^\dagger=\frac{1}{\sqrt{2}}(d^\dagger_{i,1,\uparrow}d^\dagger_{i,2,\downarrow}
-d^\dagger_{i,1,\downarrow}d^\dagger_{i,2,\uparrow})$ creates a spin
singlet pair between Cu sites on the $i$th rung.  In the Luther-Emery
universality class, charge and pairing correlations decay as power
laws in the long distance limit, with asymptotic behavior $C(r)\sim
r^{-K_\rho}$ and $P(r)\sim r^{-1/K_\rho}$, respectively.  While true
long-range superconducting order is absent in a one-dimensional
system, for $K_\rho>1$ pair correlation decay with distance is slower
than that of charge correlation and there is quasi-long range
superconducting order. Conversely, for $K_\rho<1$ charge correlations
dominate over superconducting quasi-long range order.
\begin{figure}[tb]
  \centerline{\resizebox{3.4in}{!}{\includegraphics{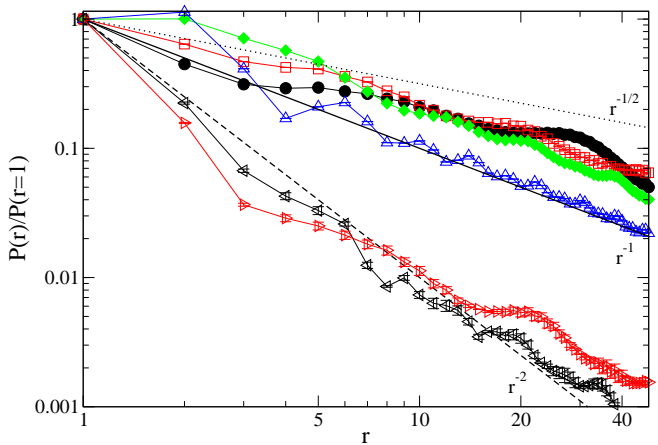}}}
  \caption{(Color online) Normalized pair-pair correlation function
    $P(r)$ as a function of the rung-rung distance $r$ for 96-rung
    ladders with $U_{\rm p}=3$ and $t_{\rm pp}=0.5$ for several
    electron dopings $\delta_e$ and hole dopings $\delta_h$.  Solid,
    dashed, and dotted lines are power laws $r^{-1}$, $r^{-2}$, and
    $r^{-1/2}$, respectively.  Circles, squares, diamonds, and up
    triangles correspond to electron dopings $\delta_e=$ 0.0625,
    0.0833, 0.125, and 0.25, respectively. Right and left triangles
    are for the hole-doped ladder with $\delta_h$ = 0.0625 and 0.125,
    respectively\cite{Song21a}. Lines are guides to the eye.}
  \label{fig2}
\end{figure}

The direct approach to determine if superconducting correlations
follow a power-law decay with distance involves fitting $P(r)$ against
$r$.  To reduce finite-size effects caused by the open boundary
conditions of our ladders \cite{Dolfi15a,Song21a}, we calculate $P(r)$
from an average of $N_{\rm avg}$ correlations of the same distance
$r$, centered about the midpoint of the ladder. The results shown here
used $N_{\rm avg}=10$ ($N_{\rm avg}=11$) for even (odd) $r$.  In
Fig.~\ref{fig2}, we show the normalized pair-pair correlation function
($P(r)/P(r=1)$) for 96-rung ladders with $U_{\rm d}=8$, $U_{\rm p}=3$,
$t_{\rm pp}=0.5$, and a range of dopings.  We find that $P(r)$ is well
fit by a power law $P(r)\sim r^{-\alpha}$ over a range of electron and
hole dopings.  As can be seen in Fig.~\ref{fig2}, there is a very
clear difference in the power law exponent for hole versus electron
doping, with a noticeably faster decay with distance for hole-doped
ladders.  For electron doping, $\alpha < 1$ over a large range of
doping, corresponding to a correlation exponent $K_\rho>1$, which
indicates quasi-long-range superconducting order.  In contrast,
$K_\rho<1$ for hole doping \cite{Song21a}. With increased hole doping,
pair correlation decays {\it faster} with distance \cite{Song21a}.

\begin{figure}[tb]
  \centerline{\resizebox{3.4in}{!}{\includegraphics{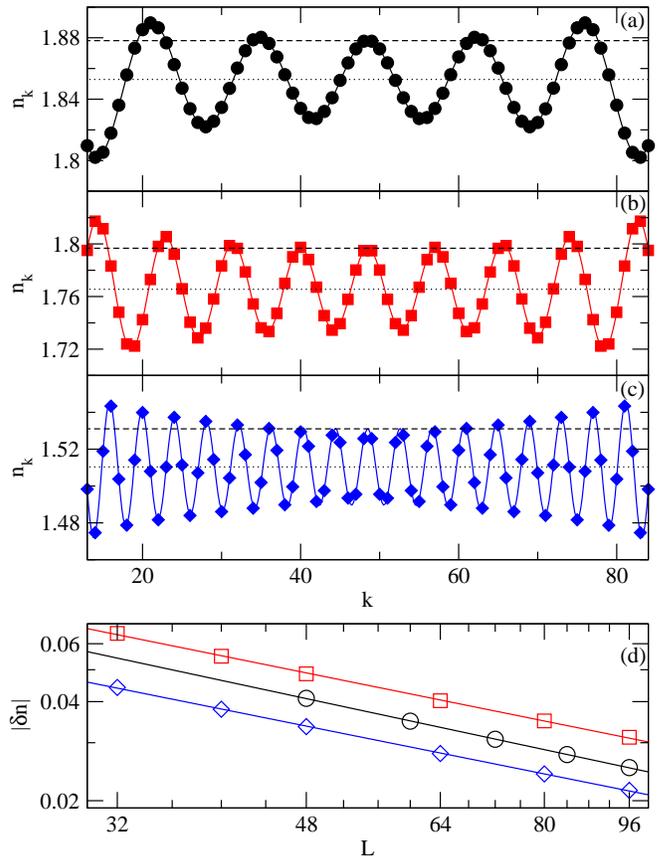}}}
  \caption{(Color online) The local charge density profile on a
    96-rung ladder with $U_{\rm p}=3$ and $t_{\rm pp}=0.5$ for
    electron dopings $\delta_e=$ (a) 0.0833, (b) 0.125, and (c) 0.25.
    The curves are fits to Eq.~\ref{friedel}.  Dotted and dashed lines
    represent $n_0$ and $n(L/2)$.  (d) Amplitude of Friedel
    oscillations at $L/2$, $\delta n$ (see text), as a function of
    ladder length $L$.  The lines are linear fits.  Circles, squares,
    and diamonds correspond to $\delta_e=$ 0.0833, 0.125, and 0.25,
    respectively.}
  \label{fig3}
\end{figure}

A more accurate approach to determining the correlation exponent
$K_\rho$ in DMRG calculations is to fit the charge density (Friedel)
oscillations caused by the open boundaries of the ladder
\cite{White02a,Dolfi15a}.  This method also permits more accurate
extrapolation of $K_\rho$ to the $L\rightarrow\infty$ limit
\cite{Dolfi15a}.  We use the following fitting function for the charge
density $n_k$ \cite{White02a,Dolfi15a,Song21a},
\begin{equation}
  n_k = n_0+ 
        A\frac{\cos(N \pi k/L_{\rm eff}+\phi)}
              {\sin(\pi k/L_{\rm eff})^{K_\rho/2}}.
  \label{friedel}
\end{equation}
In Eq.~\ref{friedel} $n_0$ is the background charge density, $A$ the
Friedel oscillation amplitude, $\phi$ a phase shift, and $L_{\rm eff}$
an effective length.  Typically $L_{\rm eff}$ is smaller than $L$ to
account for end effects \cite{Dolfi15a}.  The amplitude of the charge
density oscillations at the center of the system, $\delta
n=n(L/2)-n_0$, scales as $L^{-K_\rho/2}$. Finite-size scaling of
$\delta n$, where the values of $n_0$ and $n(L/2)$ are determined from
the fitted function in Eq.~\ref{friedel}, then yields the most precise
estimates for the correlation exponent $K_\rho$ in the infinite-length
limit ($L\rightarrow\infty$) \cite{Dolfi15a}.

In Figs.~\ref{fig3}(a)-(c) we show the Friedel oscillations of local
charge density $n_k$ on a 96-rung ladder with $U_{\rm d}=8$, $U_{\rm
  p}=3$, and $t_{\rm pp}=0.5$ for three different values of electron
doping ($\delta_e=$0.0833, 0.125, and 0.25).  For each doping level we
also provide estimates for both $n_0$ and $n(L/2)$ in
Figs.~\ref{fig3}(a)-(c).  As expected, the wavelength of the Friedel
oscillations is reduced with increasing doping $\delta_e$.  In
Fig.~\ref{fig3}(d) we show the finite-size scaling analysis for
different ladder lengths of up to $L=96$ to determine the correlation
exponent $K_\rho$ in the $L\rightarrow\infty$ limit.

\begin{figure}[tb]
  \centerline{\resizebox{3.4in}{!}{\includegraphics{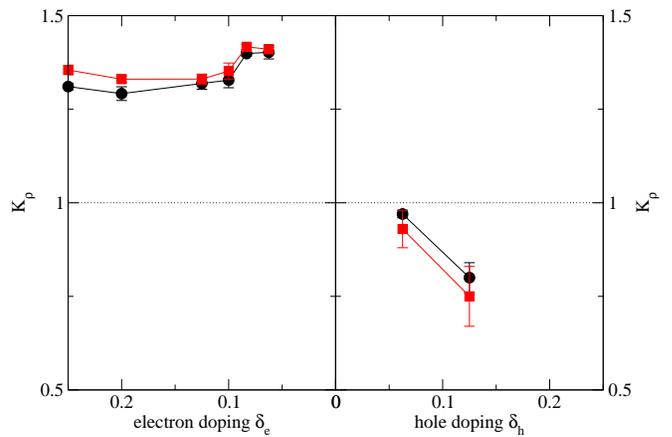}}}
  \caption{(Color online) The doping dependence of the extrapolated
    power-law exponents $K_\rho$.  Circles and squares are for
    $(U_{\rm p},t_{\rm pp})=(3,0.5)$ and $(4,0.6)$,
    respectively. Error bars are estimated from the fits in
      Fig.~\ref{fig3}(c). Lines are guides to the eye.}
  \label{fig4}
\end{figure}

In Fig.~\ref{fig4} we summarize the extrapolated values of $K_\rho$
for two sets of parameters most relevant to cuprates in both hole- and
electron-doped systems.  The values of $K_\rho$ for hole doping are
from Ref.~\onlinecite{Song21a}.  We find that for electron doping,
$K_\rho>1$ and $K_\rho$ remains nearly constant over a wide doping
range.  In contrast, for hole doping $K_\rho$ is close to 1 for very
small $\delta_h$, but rapidly decreases with $\delta_h$ and is
significantly less than 1 for $\delta_h>0.0625$. These results, 
consistent with calculations of pair-binding energies, show
that a superconducting Luther-Emery phase occurs in the electron-doped
cuprate ladder but not the hole doped ladder.

The most important conclusion from our work is that the doping
asymmetry in pairing correlations found within the one-band model
calculations for the 2D layer \cite{Jiang21a,Jiang22a,Jiang23a} occurs
also within the two-leg three-band cuprate ladder Hamiltonian for
realistic Hubbard and hopping parameters.  As in the one-band ladder,
the three-band two-leg ladder also contains rung-based spin singlets,
now on Cu-O-Cu rungs, as evidenced from the large $\Delta_s$ in the
undoped ladder (Fig.~1(a)).  Doping with electrons therefore generates
Cu$^{2+}$ ion pairs on the rungs, and superconducting correlations
persist for the same reason as in the one-band model. Doped holes
create O$^{1-}$ ions on rung or leg O-sites with equal
probability. Even when a doped hole occupies a rung O-ion, a second
doped hole necessarily occupies a neighboring leg oxygen, which cannot
be associated with any specific rung. This severely reduces the
hole-hole binding energy leading to fast decrease of the spin gap
(Fig.~1(a)).  Direct O-O hopping $t_{pp}$ is strongly pair-breaking,
as is indeed found from our calculations. This particular result has
strong implications for the 2D lattice, where individual O-atoms also
cannot be associated with any single Cu$^{2+}$-ion and each O-atom is
coupled to four other oxygens. The pair-breaking effect due to O-O
hopping therefore remains strong in 2D: absence of pairing in the
hole-doped three-band ladder necessarily implies the same for 2D. With
hindsight, this breakdown of the Zhang-Rice reduction of the full
three-band Hamiltonian to a single-band Hubbard Hamiltonian is to be
anticipated, as the original derivation by Zhang and Rice had excluded
O-O hopping \cite{Zhang88a}.

Superconductivity with electron-doping within the three-band ladder
similarly predicts the same in 2D within the three-band
Hamiltonian. Electron-doping generates spinless Cu$^{2+}$ ions in the
background antiferromagnet now instead of a spin-singlet ground state
as O$^{2-}$ ions remain closed-shell.  O-O hopping thus plays no role
whatsoever, and Cu$^{2+}$-Cu$^{2+}$ pairing, as found within the
one-band Hamiltonian will persist within the three-band
Hamiltonian. Coexistence with long-range antiferromagnetism (AFM),
as is found in the one-band calculations, is a
necessary condition of such pairing.  Such coexistence with long-range
AFM is precluded experimentally from inelastic neutron scattering
studies \cite{Motoyama07a} and muon spin rotation measurements
\cite{Saadaoui15a}.  Additionally, coexistence with AFM leads to
coupled $d_{x^2-y^2}$ and triplet pairing, as has indeed been found
within both $t$-$t^\prime$-$J$ and
$t$-$t^\prime$-$t^{\prime\prime}$-$J$ and Hubbard model calculations
\cite{Jiang21a,Jiang22a,Foley19a} also in contradiction to
experiments. We note that a recent extended $t$-$J$ model DMRG
calculation on 4- and 6-leg cylinders found dominant pairing
correlations and exponentially decaying spin correlations for electron
doping \cite{Gong21a}.  Because even-leg cylinders are expected to
possess spin gaps, distinguishing between long-range AFM and spin gap
behavior is however difficult in DMRG calculations, and these results
do not necessarily contradict those obtained in references
\cite{Jiang21a,Jiang22a} or here.

Rather than asymmetry, recent experiments have revealed deep
underlying symmetry between hole- and electron doped cuprates
\cite{Proust19a,Greene20a}.  In both cases there is absence of
coexistence between long-range AFM and SC, and there exists a quantum
critical point with Fermi surface reconstruction inside the
superconducting dome, accompanied by a sudden change in the number of
charge carriers.  In both hole- and electron-doped compounds the
carrier density is linear in doping $p$ for small doping, but jumps to
$1+p$ and $1-p$ respectively following the Fermi surface
reconstruction. The quantum critical point in hole-doped systems
occurs at the doping concentration $p_c$ where the pseudogap vanishes
at zero temperature.  The region between this critical doping and the
doping at which SC ends in both cases is occupied by a strange metal
that exhibits resistivity linear in temperature T and
magnetoresistance linear in magnetic field H
\cite{Proust19a,Greene20a,Legros19a}.  Similar behavior has now been
observed in many different families of unconventional superconductors
\cite{Doiron-Leyraud09a,Hayes21a,Yuan22a,Phillips22a}.  Many authors
have therefore speculated that there is an intimate relationship
between the quantum criticality and superconductivity.  Very recent
research indicates that charge carriers in the strange metallic state
of YBa$_2$Cu$_3$O$_7$ may be charge 2e bosons \cite{Yang22a}.  All the
above continue to be challenging within standard models of cuprate SC.

We end this Letter by pointing out that the quantum criticality and
associated phenomena can be qualitatively understood within a valence
transition theory of cuprates we have recently proposed
\cite{Mazumdar18a,Mazumdar20a,Song22a}. Within this theory the Fermi
surface reconstruction in both hole and electron-doped compounds is
due to dopant-induced transition from positive to negative charge
transfer gap state. The transition involves change in Cu-ion ionicity
from Cu$^{2+}$ to Cu$^{1+}$, resulting in transfer of nearly all
Cu-ion $d_{x^2-y^2}$ holes to the O-ions.  Similar quantum critical
transitions between different ionicities have been widely discussed
over four decades in the context of neutral-ionic transition in
organic donor-acceptor charge-transfer solids \cite{Masino17a} and
heavy fermion systems \cite{Miyake17a}.  Carrier densities of $1+p$
and $1-p$ holes are naturally expected within this approach
following the valence transition.
Transport in the normal and superconducting states with both hole- and
electron doping then involve the nearly $\frac{3}{4}$-filled strongly
correlated O-band alone, explaining the mysterious symmetry between
the two cases.  Previous calculations on the single-band 2D
$\frac{3}{4}$-filled Hubbard Hamiltonian have shown that (a) precisely
at this carrier concentration there is a strong tendency to transition
to a paired-electron crystal (PEC), which is a charge-ordered state of
spin-singlet electron pairs \cite{Li10a,Dayal11a}, and (b) very close
to this concentration there occurs enhancement of superconducting pair
correlations by the Hubbard $U$ \cite{Gomes16a,Clay19a}. In the
absence of phase coherence the spin-coupled electron pairs can
conceivably be the bosonic charge carriers in the strange metallic
state. Importantly, the occurrence of the strange metallic state under
pressure in the organic superconductor (TMTSF)$_2$PF$_6$
\cite{Doiron-Leyraud09a}, known to possess a $\frac{1}{4}$-filled hole
band ($\frac{3}{4}$-filled electron band) is indirect confirmation of
this approach.  These and related topics are currently under
investigation.

Work at Arizona was supported by National Science Foundation (NSF)
grant NSF-CHE-1764152.  Some of the calculations were performed using
high performance computing resources maintained by the University of
Arizona Research Technologies department and supported by the University of
Arizona Technology and Research Initiative Fund (TRIF), University
Information Technology Services (UITS), and Research, Innovation, and
Impact (RII).

\end{document}